\documentclass[preprint,prd,aps,nofootinbib]{revtex4}
\usepackage{graphicx}
\usepackage{color}
\usepackage{CJK,upgreek,fancyhdr}
\usepackage{bm}

\UseRawInputEncoding

\begin{document}

\begin{CJK*}{GBK}{song}


\title{Neutrinos in the flavor-dependent $U(1)_F$ model}

\author{Jin-Lei Yang$^{1,2,3}$\footnote{jlyang@hbu.edu.cn},Jie Li$^{1,2,3}$\footnote{jennyli305@163.com}}

\affiliation{Department of Physics, Hebei University, Baoding, 071002, China$^1$\\
Key Laboratory of High-precision Computation and Application of Quantum Field Theory of Hebei Province, Baoding, 071002, China$^2$\\
Research Center for Computational Physics of Hebei Province, Baoding, 071002, China$^3$}

\begin{abstract}
The neutrino oscillation experiments provide definitive evidence of new physics beyond the Standard Model (SM), and the neutrino mass-squared differences and flavor mixing have been precisely measured. This study examines the neutrino sector within the flavor-dependent $U(1)_F$ model, where the unique fermion sector can simultaneously address both the flavor mixing puzzle and the mass hierarchy puzzle. It is found that the lightest neutrino is naturally massless in this model, and the predicted neutrino mass-squared differences, flavor mixing angles, Dirac CP phase agree well with the experimental measurements. Additionally, the effects of the Dirac CP phase and Majorana CP phase on the theoretical predictions of the neutrino transition magnetic dipole moments are analyzed.

\end{abstract}

\maketitle

\section{Introduction\label{sec1}}

The fermion sector in the Standard Model (SM) exhibits a large hierarchical structure of masses across the three families, and the quark flavor mixings described by the Cabibbo-Kobayashi-Maskawa (CKM) matrix~\cite{Cabibbo:1963yz,Kobayashi:1973fv} are not predicted from first principles within the SM. These phenomena are known as the mass hierarchy puzzle and the flavor puzzle, which indicate the observed fermionic mass spectrum and mixings are still enigmatic in particle physics. Additionally, the observed tiny but nonzero neutrino masses and flavor mixing among the three generations of neutrinos not only make the flavor puzzle in the SM more acutely, but also provide unambiguous evidence of new physics (NP) beyond the SM.

Addressing the mass hierarchy puzzle, flavor puzzle, nonzero neutrino masses and neutrino mixings, the flavor-dependent $U(1)_F$ model (FDM) proposed in our previous work~\cite{Yang:2024kfs,Yang:2024znv} simultaneously accounts for these phenomena. Furthermore, the analysis demonstrates that the six quark masses, the CKM matrix can be accurately fitted within the model. This study focuses on exploring whether the neutrino sector within the FDM can account for the neutrino-related observations. Experimentally, the observed neutrino mass-squared differences are~\cite{ParticleDataGroup:2024cfk}
\begin{eqnarray}
&&\Delta m_{12}^2\equiv m_{\nu_2}^2-m_{\nu_1}^2=(7.53\pm0.18)\times 10^{-5}\;{\rm eV}^2,\nonumber\\
&&\Delta m_{23}^2\equiv m_{\nu_3}^2-m_{\nu_2}^2\approx(2.455\pm0.028)\times 10^{-3}\;{\rm eV}^2,\;({\rm NH})\nonumber\\
&&\Delta m_{32}^2\equiv m_{\nu_2}^2-m_{\nu_3}^2\approx(2.529\pm0.029)\times 10^{-3}\;{\rm eV}^2,\;({\rm IH}) \label{eq1}
\end{eqnarray}
where NH, IH correspond to normal hierarchy (NH) and inverse hierarchy (IH) neutrino masses respectively. For Majorana neutrinos, the observed neutrino oscillations can be expressed as
\begin{eqnarray}
&&U_{\rm PMNS}=\left(\begin{array}{ccc} c_{12}c_{13} & s_{12}c_{13} & s_{13}e^{-i\delta}\\
-s_{12}c_{23}-c_{12}s_{13}s_{23}e^{i\delta} & c_{12}c_{23}-s_{12}s_{13}s_{23}e^{i\delta} & c_{13}s_{23},\\
s_{12}s_{23}-c_{12}s_{13}c_{23}e^{i\delta} & -c_{12}s_{23}-s_{12}s_{13}c_{23}e^{i\delta} & c_{13}c_{23}\end{array}\right)\nonumber\\
&&\qquad\qquad\times{\rm diag}(e^{i\rho},e^{i\sigma},1), \label{eq2}
\end{eqnarray}
where $c_{ij}\equiv \cos\theta_{ij}$, $s_{ij}\equiv \sin\theta_{ij}$, $\delta$ is usually referred to as the Dirac CP phase, and $\rho,\;\sigma$ are the possible CP phases for Majorana neutrinos. Particle Data Group (PDG)~\cite{ParticleDataGroup:2024cfk} collects the measured results of the mixing angles $\theta_{ij}$ and the Dirac CP phase $\delta$, the results read
\begin{eqnarray}
&&s_{12}^2=0.307\pm0.013,\;s_{13}^2=(2.19\pm0.07)\times 10^{-2},\;\delta=(1.19\pm0.22)\pi\nonumber\\
&&s_{23}^2=0.558^{+0.015}_{-0.021}\;\;({\rm NH}),\;s_{23}^2=0.553^{+0.016}_{-0.024}\;\;({\rm IH}). \label{eq3}
\end{eqnarray}

Massive and mixing neutrinos can induce the nonzero neutrino transition magnetic dipole moments (MDM), which are predicted to be zero in the SM. Furthermore, Dirac and Majorana neutrinos exhibit different structures in their transition MDM~\cite{Fujikawa:1980yx,Pal:1981rm,Nieves:1981zt,Kayser:1982br,Shrock:1982sc,Kayser:1984ge,Bell:2005kz,Bell:2006wi,Giunti:2014ixa,Xu:2019dxe,Ge:2022cib}. Therefore, observing the neutrino transition MDM is a crucial method to explore the neutrino-related NP. The transition MDM can be tested experimentally in various ways. For instance, measurements of neutrino scattering cross sections with electron peaks in the low momentum transfer region (e.g., the reactor experiment GEMMA~\cite{Beda:2013mta} and the solar experiment Borexino~\cite{Borexino:2017fbd}), stellar cooling observations of red giants~\cite{Diaz:2019kim} and white dwarfs~\cite{Corsico:2014mpa,MillerBertolami:2014oki,Hansen:2015lqa}, and future dark matter direct detection experiments such as Xenon1T~\cite{XENON:2020rca} and PandaX~\cite{PandaX-II:2020udv,PandaX-II:2021nsg} hold the potential to observe the neutrino transition MDM. Additionally, nonzero Majorana neutrino transition MDM within the core of supernova explosions may leave a potentially observable imprint on the energy spectra of neutrinos and antineutrinos from supernovae~\cite{deGouvea:2012hg,deGouvea:2013zp}. Therefore, we also present the predicted neutrino transition MDM in the FDM.

The paper is organized as follows: In Sec.\ref{sec2}, we present the lepton sector, including the charged lepton mass matrix, the neutrino mass matrix, and the analytical calculations of the neutrino transition MDM within the FDM. In Sec.\ref{sec3}, we first investigate whether the lepton sector within the FDM can accommodate the measured neutrino mass-squared differences and flavor mixings. Subsequently, the numerical results of the neutrino transition MDM are presented and analyzed. We conclude the paper with a summary of findings and discussions in Sec.~\ref{sec4}.

\section{Lepton sector and neutrino transition MDM in the FDM\label{sec2}}

The FDM extends the SM by a $U(1)_F$ local gauge group associated with the particles' flavor, with the third-generation fermions possessing zero $U(1)_F$ charge. Consequently, the third generation of right-handed neutrino is trivial under $SU(2)_L\otimes U(1)_Y\otimes U(1)_F$. As a result, only two right-handed neutrinos with nonzero $U(1)_F$ charges are introduced, and naturally realizing the so-called minimal see-saw~\cite{Ma:1998zg,Frampton:2002qc,Xing:2020ald} in this model\footnote{The minimal seesaw mechanism warrants consideration and study for several compelling reasons, for example, its predictive power is enhanced by the significant reduction in the number of free parameters associated with the minimal seesaw mechanism. Further analysis about the minimal see-saw can be found in Refs.~\cite{Gu:2006wj,Davidson:2006tg,Chan:2007ng,Guo:2006qa,Ren:2008yi,Xing:2007zj,Xing:2009ce,Hirsch:2009ra,Deppisch:2010fr,Xing:2011ur,Mondal:2012jv,Abada:2014vea,Abada:2014zra,
Luo:2014upa,Mohapatra:2015gwa,Nath:2018hjx,Huang:2018wqp,Xing:2020ezi,CarcamoHernandez:2019eme}.}. The charges of fermions corresponding to $SU(2)_L\otimes U(1)_Y\otimes U(1)_F$ are
\begin{eqnarray}
&&L_{1}\sim(2,Y_{L},z),L_{2}\sim(2,Y_{L},-z),L_{3}\sim(2,Y_{L},0),\nonumber\\
&&R_{1}\sim(1,Y_{R},-z),R_{2}\sim(1,Y_{R},z),R_{3}\sim(1,Y_{R},0),\nonumber\\
&&\nu_{R_1}\sim(1,0,-z),\nu_{R_2}\sim(1,0,z),\label{eq4}
\end{eqnarray}
where nonzero $z$ denotes the $U(1)_F$ charge, $L_i$ ($L=l,\;q$) denote the $i-$generation left-handed fermion doublets, $R_{i}$ ($R=u,\;d,\;e$) denotes the $i-$generation right-handed fermion singlets in the SM, $\nu_{R_i}\;(i=1,2)$ denote the right-handed neutrinos. The scalar sector of the model is extended by two doublets and one singlet
\begin{eqnarray}
&&\Phi_1=\left(\begin{array}{c}\phi_1^+\\ \frac{1}{\sqrt2}(i A_1+S_1+v_1)\end{array}\right)\sim(2,\frac{1}{2},z),\nonumber\\
&&\Phi_2=\left(\begin{array}{c}\phi_2^+\\ \frac{1}{\sqrt2}(i A_2+S_2+v_2)\end{array}\right)\sim(2,\frac{1}{2},-z),\nonumber\\
&&\Phi_3=\left(\begin{array}{c}\phi_3^+\\ \frac{1}{\sqrt2}(i A_3+S_3+v_3)\end{array}\right)\sim(2,\frac{1}{2},0),\nonumber\\
&&\chi=\frac{1}{\sqrt2}(i A_{\chi}+S_{\chi}+v_\chi)\sim(1,0,2z),\label{eq5}
\end{eqnarray}
where $v_i\;(i=1,\;2,\;3),\;v_\chi$ are the VEVs of $\Phi_i,\;\chi$ respectively.

Based on the local gauge symmetry and gauge charges in Eqs.~(\ref{eq4}, \ref{eq5}), the Yukawa couplings in the FDM can be written as
\begin{eqnarray}
&&\mathcal{L}_Y=Y_u^{33}\bar q_3 \tilde \Phi_3 u_{R_3}+Y_d^{33}\bar q_3 \Phi_3 d_{R_3}+Y_u^{32}\bar q_3 \tilde{\Phi}_1 u_{R_2}+Y_u^{23}\bar q_2 \tilde \Phi_1 u_{R_3}+Y_d^{32}\bar q_3 \Phi_2 d_{R_2}\nonumber\\
&&\qquad\; +Y_d^{23}\bar q_2 \Phi_2 d_{R_3}+Y_u^{21}\bar q_2 \tilde{\Phi}_3 u_{R_1}+Y_u^{12}\bar q_1 \tilde \Phi_3 u_{R_2}+Y_d^{21}\bar q_2 \Phi_3 d_{R_1}+Y_d^{12}\bar q_1 \Phi_3 d_{R_2}\nonumber\\
&&\qquad\;+Y_u^{31}\bar q_3 \tilde{\Phi}_2 u_{R_1}+Y_u^{13}\bar q_1 \tilde \Phi_2 u_{R_3}+Y_d^{31}\bar q_3 \Phi_1 d_{R_1}+Y_d^{13}\bar q_1 \Phi_1 d_{R_3}+Y_e^{33}\bar l_3 \Phi_3 e_{R_3}\nonumber\\
&&\qquad\;+Y_e^{32}\bar l_3 \Phi_2 e_{R_2}+Y_e^{23}\bar l_2 \Phi_2 e_{R_3}+Y_e^{21}\bar l_2 \Phi_3 e_{R_1}+ Y_e^{12}\bar l_1 \Phi_3 e_{R_2}+Y_e^{31}\bar l_3 \Phi_1 e_{R_1}\nonumber\\
&&\qquad\;+Y_e^{13}\bar l_1 \Phi_1 e_{R_3}+Y_R^{11}\bar\nu^c_{R_1}\nu_{R_1}\chi+Y_R^{22}\bar\nu^c_{R_2}\nu_{R_2} \chi^*+Y_D^{21}\bar l_2 \tilde \Phi_3 \nu_{R_1}+Y_D^{12}\bar l_1 \tilde \Phi_3 \nu_{R_2}\nonumber\\
&&\qquad\;+Y_D^{31}\bar l_3 \tilde \Phi_2 \nu_{R_1}+Y_D^{32}\bar l_3 \tilde \Phi_1 \nu_{R_2}+h.c..\label{eq6}
\end{eqnarray}
As mentioned above, the quark sector in this model have been analyzed in our previous work~\cite{Yang:2024kfs,Yang:2024znv}, we focus on the lepton sector in this work. After $\Phi_1,\;\Phi_2,\;\Phi_3,\;\chi$ receive nonzero VEVs, the mass matrices of charged leptons and neutrinos can be written as
\begin{eqnarray}
&&m_e=\left(\begin{array}{ccc} 0 & m_{e,12} & m_{e,13}\\
m_{e,21} & 0 & m_{e,23}\\
m_{e,31} & m_{e,32} & m_{e,33}\end{array}\right),\;m_\nu=\left(\begin{array}{cc} 0 & M_D^T\\
M_D & M_R\end{array}\right),\label{eq7}
\end{eqnarray}
where $M_D$ is $2\times3$ Dirac mass matrix and $M_R$ is $2\times2$ Majorana mass matrix, and
\begin{eqnarray}
&&m_{e,11}=m_{e,22}=0,\;m_{e,33}=\frac{1}{\sqrt2}Y_e^{33}v_3,\;m_{e,12}=\frac{1}{\sqrt2}Y_e^{12}v_3,\;m_{e,21}=\frac{1}{\sqrt2}Y_e^{21}v_3,\nonumber\\
&&m_{e,13}=\frac{1}{\sqrt2}Y_e^{13}v_1,\;m_{e,31}=\frac{1}{\sqrt2}Y_e^{31}v_1,\;m_{e,23}=\frac{1}{\sqrt2}Y_e^{23}v_2,\;m_{e,32}=\frac{1}{\sqrt2}Y_e^{32}v_2,\nonumber\\
&&M_{D,11}=M_{D,22}=0,\;M_{D,12}=\frac{1}{\sqrt2}Y_D^{12}v_3,\;M_{D,21}=\frac{1}{\sqrt2}Y_D^{21}v_3,\;M_{D,31}=\frac{1}{\sqrt2}Y_D^{31}v_1,\nonumber\\
&&M_{D,32}=\frac{1}{\sqrt2}Y_D^{32}v_2,\;M_{R,12}=M_{R,21}=0,\;M_{R,11}=\frac{1}{\sqrt2}Y_R^{11}v_\chi,\;M_{R,22}=\frac{1}{\sqrt2}Y_R^{22}v_\chi.\label{eq8}
\end{eqnarray}
The Hermitian charged lepton mass matrix\footnote{The Hermitian property ensures that the eigenvalues of charged lepton mass matrix are real, which represent the observed masses of the charged leptons. Additionally, the invariance under charge conjugation in the lepton sector further supports the requirement for the mass matrix to be Hermitian.} requires real $m_{e,33}$ and
\begin{eqnarray}
&&Y_e^{12}=Y_e^{21*},\;Y_e^{13}=Y_e^{31*},\;Y_e^{23}=Y_e^{32*}.
\end{eqnarray}
Then the Hermitian mass matrix for charged lepton can be written as
\begin{eqnarray}
&&m_e=\left(\begin{array}{ccc} 0 & m_{e,12} & m_{e,13}\\
m_{e,12}^* & 0 & m_{e,23}\\
m_{e,31}^* & m_{e,32}^* & m_{e,33}\end{array}\right).\label{eq10}
\end{eqnarray}

Taking the measured charged lepton masses as inputs, one can obtain $m_{e,33}$, $|m_{e,23}|$, $|m_{e,12}|$ under the approximation $|m_{e,12}|,\;|m_{e,13}|,\;|m_{e,23}|\ll m_{e,33}$ as
\begin{eqnarray}
&&m_{e,33}=m_{e_1}+m_{e_2}+m_{e_3},\nonumber\\
&&|m_{e,23}|=[(m_{e_1}+m_{e_2})m_{e,33}-|m_{e,13}|^2]^{1/2},\nonumber\\
&&|m_{e,12}|=\frac{|m_{e,13}||m_{e,23}|}{|m_{e,33}|} \cos(\theta_{e,12}+\theta_{e,23}-\theta_{e,13})\nonumber\\
&&\qquad\qquad+\Big\{[\frac{|m_{e,13}||m_{e,23}|}{|m_{e,33}|^2}\cos(\theta_{e,12}+\theta_{e,23}-\theta_{e,13})]^2+\frac{m_{e_1}m_{e_2}}{|m_{e,33}|^2}\Big\}^{1/2}|m_{e,33}|,\label{eq9}
\end{eqnarray}
where $\theta_{e,ij}\;(ij=12,\;13,\;23)$ are defined as $m_{e,ij}=|m_{e,ij}|e^{i\theta_{e,ij}}$, and $m_{e_k}\;(k=1,\;2,\;3)$ is $k-$generation charged lepton mass. For the $5\times5$ neutrino mass matrix $m_\nu$ defined in Eq.~(\ref{eq7}), we can derive the $3\times3$ light Majorana neutrino mass matrix $\hat m_\nu$ by taking $M_D^TM_R^{-1}\ll 1$
\begin{eqnarray}
&&\hat m_\nu\approx -M_D^TM_R^{-1} M_D=-\left(\begin{array}{ccc} \frac{M_{D,12}^2}{M_{R,22}} & 0 & \frac{M_{D,12}M_{D,32}}{M_{R,22}}\\
0 & \frac{M_{D,21}^2}{M_{R,11}} & \frac{M_{D,21}M_{D,31}}{M_{R,11}}\\
\frac{M_{D,12}M_{D,32}}{M_{R,22}} & \frac{M_{D,21}M_{D,31}}{M_{R,11}} & \frac{M_{D,31}^2}{M_{R,11}}+\frac{M_{D,32}^2}{M_{R,22}}\end{array}\right).
\end{eqnarray}
$\hat m_\nu$ can be simplified by simple calculation and redefined as
\begin{eqnarray}
&&\hat m_\nu=\left(\begin{array}{ccc} \hat m_{\nu,11} & 0 & \hat m_{\nu,13}\\
0 & \hat m_{\nu,22} & \hat m_{\nu,23}\\
\hat m_{\nu,13} & \hat m_{\nu,23} & \frac{\hat m_{\nu,13}^2}{\hat m_{\nu,11}}+\frac{\hat m_{\nu,23}^2}{\hat m_{\nu,22}}\end{array}\right),\label{eq11}
\end{eqnarray}
where
\begin{eqnarray}
&&\hat m_{\nu,11}=-\frac{M_{D,12}^2}{M_{R,22}},\;\hat m_{\nu,22}=-\frac{M_{D,21}^2}{M_{R,11}},\nonumber\\
&&\hat m_{\nu,13}=-\frac{M_{D,12}M_{D,32}}{M_{R,22}},\;\hat m_{\nu,23}=-\frac{M_{D,21}M_{D,31}}{M_{R,11}}.
\end{eqnarray}

It is straightforward to verify that the determinant of the mass matrix defined in Eq.~(\ref{eq11}) is zero, indicating that one of the three light neutrinos in the FDM must be massless. In this scenario, the masses of the other two light neutrinos can be fixed by the measured neutrino mass-squared differences in Eq.~(\ref{eq1}), i.e. for the NH and IH neutrino masses we have
\begin{eqnarray}
&&m_{\nu_1}^{\rm NH}=0\;{\rm eV},\;m_{\nu_2}^{\rm NH}=(0.753\pm0.018)^{1/2}\times 10^{-2}\;{\rm eV},\nonumber\\
&&m_{\nu_3}^{\rm NH}=(25.303\pm0.298)^{1/2}\times 10^{-2}\;{\rm eV},\\
&&m_{\nu_3}^{\rm IH}=0\;{\rm eV},\;m_{\nu_1}^{\rm IH}=(24.537\pm0.308)^{1/2}\times 10^{-2}\;{\rm eV},\nonumber\\
&&m_{\nu_2}^{\rm IH}=(25.29\pm0.29)^{1/2}\times 10^{-2}\;{\rm eV}.
\end{eqnarray}
Taking the neutrino masses above as inputs, we can fix two of the free parameters in Eq.~(\ref{eq11}) by solving the two following equations
\begin{eqnarray}
&&\Big[|\hat m_{\nu,22}|^2(|\hat m_{\nu,11}|^2+|\hat m_{\nu,13}|^2-|\hat m_{\nu,11}||\hat m_{\nu,22}|)^2+2|\hat m_{\nu,11}|^2|\hat m_{\nu,22}||\hat m_{\nu,23}|^2(|\hat m_{\nu,22}|\nonumber\\
&&-|\hat m_{\nu,11}|)+|\hat m_{\nu,11}|^2\times|\hat m_{\nu,23}|^4+2|\hat m_{\nu,11}||\hat m_{\nu,22}||\hat m_{\nu,13}|^2|\hat m_{\nu,23}|^2\cos\theta\Big]\nonumber\\
&&\times\Big[|\hat m_{\nu,22}|^2(|\hat m_{\nu,13}|^2+|\hat m_{\nu,11}|^2+|\hat m_{\nu,11}||\hat m_{\nu,22}|)^2+2|\hat m_{\nu,11}|^2 |\hat m_{\nu,23}|^2|\hat m_{\nu,22}|(|\hat m_{\nu,11}|\nonumber\\
&&+|\hat m_{\nu,22}|)+|\hat m_{\nu,11}|^2|\hat m_{\nu,23}|^4+2|\hat m_{\nu,13}|^2|\hat m_{\nu,23}|^2|\hat m_{\nu,11}||\hat m_{\nu,22}|\cos\theta\Big]\nonumber\\
&&=|m_{\nu,11}|^4|m_{\nu,22}|^4(m_{\nu_a}^2-m_{\nu_b}^2)^2,\label{eq15}\\
&&|\hat m_{\nu,11}|^4|\hat m_{\nu,22}|^2+|\hat m_{\nu,11}|^2|\hat m_{\nu,22}|^4+2|\hat m_{\nu,13}|^2|\hat m_{\nu,11}|^2|\hat m_{\nu,22}|^2+2|\hat m_{\nu,23}|^2|\hat m_{\nu,11}|^2|\hat m_{\nu,22}|^2\nonumber\\
&&+|\hat m_{\nu,13}|^4|\hat m_{\nu,22}|^2+|\hat m_{\nu,23}|^4|\hat m_{\nu,11}|^2+2|\hat m_{\nu,13}|^2|\hat m_{\nu,23}|^2|\hat m_{\nu,11}||\hat m_{\nu,2}|\cos\theta\nonumber\\
&&=|\hat m_{\nu,11}|^2|\hat m_{\nu,22}|^2(m_{\nu_a}^2+m_{\nu_b}^2),\label{eq16}
\end{eqnarray}
where $\theta=2\theta_{\nu,13}+\theta_{\nu,22}-\theta_{\nu,11}-2\theta_{\nu,23}$, and $\theta_{\nu,\alpha\beta}$ ($\alpha\beta=11,\;22,\;13,\;23$) is the phase of $\hat m_{\nu,\alpha\beta}$, i.e. $\hat m_{\nu,\alpha\beta}=|\hat m_{\nu,\alpha\beta}|e^{i\theta_{\nu,\alpha\beta}}$.

Considering the nonzero neutrino masses and flavor mixings, we can obtain the nonzero neutrino transition MDM, which may have significant astrophysical consequences. Generally, the MDM of Dirac fermions can be written as
\begin{eqnarray}
&&\mathcal{L}_{\rm MDM}=\frac{1}{2}\mu_{ij}^D\bar\psi_i\sigma^{\mu\nu}\psi_jF_{\mu\nu},
\end{eqnarray}
where $i,j$ are the indices of generation, $\psi_{i,j}$ are the four component fermions, $F_{\mu\nu}$ is the electromagnetic field strength, $\sigma^{\mu\nu}=\frac{i}{2}[\gamma^\mu,\gamma^\nu]$, and $\mu_{ij}^D$ ($i\neq j$) is the transition MDM between $\psi_i$ and $\psi_j$. After matching between the effective theory and the full theory, one can get all high dimension operators together with their coefficients. And it is enough to retain the dimension 6 operators~\cite{Feng:2008cn,Feng:2008nm,Feng:2009gn}
\begin{eqnarray}
&&{\mathcal O}_1^{L,R}=e\bar \psi_i(i/\!\!\!\!\mathcal{D})^3P_{L,R}\psi_j,\nonumber\\
&&{\mathcal O}_2^{L,R}=e\overline{(i{\mathcal D}_\mu \psi_i)}\gamma^\mu F\cdot\sigma P_{L,R}\psi_j,\nonumber\\
&&{\mathcal O}_3^{L,R}=e\bar \psi_i F\cdot\sigma\gamma^\mu P_{L,R}(i{\mathcal D}_\mu \psi_j),\nonumber\\
&&{\mathcal O}_4^{L,R}=e\bar \psi_i(\partial^\mu F_{\mu\nu})\gamma^\nu P_{L,R}\psi_j,\nonumber\\
&&{\mathcal O}_5^{L,R}=em_{\psi_i}\bar \psi_i(i/\!\!\!\!\mathcal{D})^2P_{L,R}\psi_j,\nonumber\\
&&{\mathcal O}_6^{L,R}=em_{\psi_i}\bar \psi_iF\cdot\sigma P_{L,R}\psi_j,\label{O6}
\end{eqnarray}
in the following calculations. In Eq.~(\ref{O6}), ${\mathcal D}_\mu=\partial_\mu+ie A_\mu$ and $A_\mu$ denotes the photon field, $P_{L,R}=(1\mp\gamma_5)/2$, $m_{\psi_i}$ is the mass of fermion $\psi_i$. It is obvious that, all dimension $6$ operators in Eq.~(\ref{O6}) induce the effective couplings among photons and fermions, and only the operators ${\mathcal O}_2^{L,R}$, ${\mathcal O}_3^{L,R}$, ${\mathcal O}_6^{L,R}$ can make contributions to the MDM of fermions. In addition, if the full theory is invariant under the combined transformation of charge conjugation, parity and time reversal (CPT), the Wilson coefficients $C_2^{L,R}$, $C_3^{L,R}$, $C_6^{L,R}$ of the corresponding operators ${\mathcal O}_2^{L,R}$, ${\mathcal O}_3^{L,R}$, ${\mathcal O}_6^{L,R}$ satisfy the relations
\begin{eqnarray}
&&C_3^{L,R}=(C_2^{R,L})^*,\;\;\;C_6^{L}=(C_6^{R})^*.
\end{eqnarray}
After applying the equations of motion to the external fermions, the transition MDM $\mu_{ij}^D$ can be written in the form of $C_2^{L}$, $C_2^{R}$, $C_6^{R}$ as
\begin{eqnarray}
\mu_{ij}^D=4m_e \Re(m_{\psi_i}C_2^{R}+m_{\psi_j}C_2^{L*}+m_{\psi_i}C_6^{R})\mu_B,
\end{eqnarray}
where $\Re(\cdot\cdot\cdot)$ denote the operation to take the real part of a complex number, $m_e$ is the electron mass, the Bohr magneton $\mu_B$ is defined as $\mu_B\equiv e/(2m_e)$.

\begin{figure}
\setlength{\unitlength}{1mm}
\centering
\includegraphics[width=4.5in]{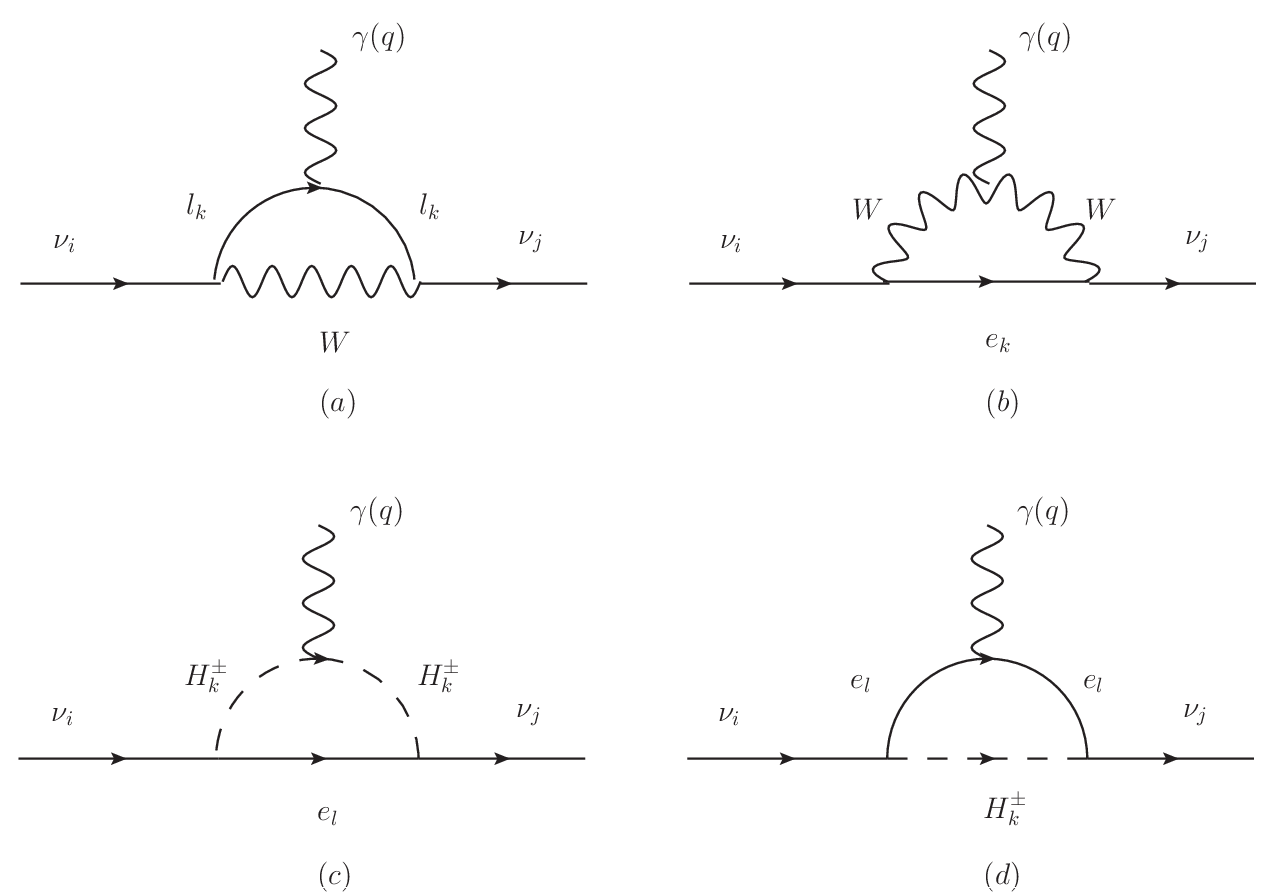}
\vspace{0cm}
\caption[]{The one-loop diagrams contributing to the TMM of Majorana neutrinos in the FDM.}
\label{Fig1}
\end{figure}

In the FDM, the Feynman diagrams contributing to neutrino transition MDM are plotted in Fig.~\ref{Fig1}. Then the nonzero coefficients can be written by neglecting the tiny Yukawa couplings of neutrinos as
\begin{eqnarray}
&&C_2^{L(a)}=\frac{1}{2M_W^2}C_L^{We_k\nu_i}C_L^{We_k\nu_j}\Big[I_1(x_{e_k},x_W)-I_3(x_{e_k},x_W)\Big],\nonumber\\
&&C_2^{L(b)}=\frac{1}{2M_W^2}C_L^{We_k\nu_i}C_L^{We_k\nu_j}\Big[I_2(x_{e_k},x_W)+I_3(x_{e_k},x_W)\Big],\nonumber\\
&&C_2^{L(c)}=\frac{1}{4M_W^2}C_L^{H_k \bar e_l \nu_i}C_L^{H_k \bar e_l \nu_j}\Big[I_3(x_{e_l},x_{H_k})-I_2(x_{e_l},x_{H_k})\Big],\nonumber\\
&&C_2^{L(d)}=\frac{1}{4M_W^2}C_L^{H_k \bar e_l \nu_i}C_L^{H_k \bar e_l \nu_j}\Big[2I_2(x_{e_l},x_{H_k})-I_1(x_{e_l},x_{H_k})-I_3(x_{e_l},x_{H_k})\Big],\label{eq20}
\end{eqnarray}
where $C_{abc}^L$ is the coupling constant for left-hand part of particles $a,\;b,\;c$, $x_i=m_i^2/M_W^2$, $M_W$ is the W boson mass and~\cite{Zhang:2014iva,Yang:2021duj}
\begin{eqnarray}
&&I_1(x_1,x_2)=\frac{1}{16\pi^2}\Big[\frac{1+\log x_2}{x_2-x_1}-\frac{x_1\log x_1-x_2 \log x_2}{(x_2-x_1)^2}\Big],\nonumber\\
&&I_2(x_1,x_2)=\frac{1}{32\pi^2}\Big[\frac{3+2\log x_2}{x_2-x_1}-\frac{2x_2+4x_2\log x_2}{(x_2-x_1)^2}-\frac{2x_1^2\log x_1-2x_2^2\log x_2}{(x_2-x_1)^3}\Big],\nonumber\\
&&I_3(x_1,x_2)=\frac{1}{96\pi^2}\Big[\frac{11+6\log x_2}{x_2-x_1}-\frac{15x_2+18x_2\log x_2}{(x_2-x_1)^2}+\frac{6x_2^2+18x_2^2\log x_2}{(x_2-x_1)^3}\nonumber\\
&&\qquad\qquad\quad+\frac{6x_1^3\log x_1-6x_2^3\log x_2}{(x_2-x_1)^4}\Big].
\end{eqnarray}
Eq.~(\ref{eq20}) shows that the contributions from the charged Higgs are proportional to the couplings $C_L^{H_k \bar e_l \nu_i}$, which are suppressed by the small charged lepton masses. As a result, the contributions to the neutrino transition MDM in the FDM are dominated by those from the $W$ boson-related interactions. Moreover, only $C_2^L=C_2^{L(a)}+C_2^{L(b)}+C_2^{L(c)}+C_2^{L(d)}$ is predicted to be nonzero in the FDM as shown in Eq.~(\ref{eq20}), hence the transition MDM of Majorana neutrinos in the FDM can be written as
\begin{eqnarray}
\mu_{ij}=\mu_{ij}^D-\mu_{ji}^D,
\end{eqnarray}
with
\begin{eqnarray}
\mu_{ij}^D=4m_e \Re(m_{\nu_j} C_2^{L*})\mu_B,
\end{eqnarray}
and $m_{\nu_j}$ denotes the neutrino mass.

XENONnT~\cite{XENON:2022ltv} presented an upper bound on the MDM of solar neutrinos, which reads $\mu_{\rm solar}<6.3\times10^{-12}\mu_B$. In general, $\mu_{\rm solar}$ can be approximated by the transition MDM of neutrinos approximately as~\cite{Akhmedov:2022txm}
\begin{eqnarray}
&&\mu_{\rm solar}=|\mu_{12}|^2c_{13}^2+|\mu_{13}|^2(c_{13}^2 c_{12}^2+s_{13}^2)+|\mu_{23}|^2(c_{13}^2 s_{12}^2+s_{13}^2).
\end{eqnarray}

\section{Numerical results\label{sec3}}

The measured neutrino mass-squared differences in Eq.~(\ref{eq1}), flavor mixings in Eq.~(\ref{eq2}) and charged lepton masses will impose strict constraints on the parameters in Eq.~(\ref{eq8}). Generally, $m_e$ in Eq.~(\ref{eq7}) and $\hat m_\nu$ in Eq.~(\ref{eq11}) can be diagonalized by
\begin{eqnarray}
&&m_e^{\rm diag}=U_L^{e,\dagger} m_e U_R^{e},\;m_\nu^{\rm diag}=U^{\nu, T}\hat m_\nu U^{\nu},
\end{eqnarray}
where $U_L^{e},\;U_R^{e},\;U^{\nu}$ are the $3\times 3$ unitary matrices. Then $U_{\rm PMNS}$ in the FDM can be defined theoretically as
\begin{eqnarray}
&&U_{\rm PMNS}=U_L^{e,\dagger}\cdot U^{\nu}.
\end{eqnarray}
PDG collects the measured results of $|U_{\rm PMNS}|$ and Dirac CP phase $\delta$ in $3\sigma$ ranges, and we consider~\cite{ParticleDataGroup:2024cfk}
\begin{eqnarray}
&&0.801<|U_{\rm PMNS}|_{11}<0.842,\;0.518<|U_{\rm PMNS}|_{12}<0.580,\;0.143<|U_{\rm PMNS}|_{13}<0.155,\nonumber\\
&&0.498<|U_{\rm PMNS}|_{22}<0.690,\;0.634<|U_{\rm PMNS}|_{23}<0.770,\;0.53\pi<\delta<1.85\pi\label{eq25}
\end{eqnarray}
in the numerical calculations. To confront the predictions of $U_{\rm PMNS}$ with the experimental data, we extract the standard-parametrization parameters according to the formulas~\cite{Xing:2020ald}
\begin{eqnarray}
&&s_{13}=|U_{\rm PMNS}|_{13},\;s_{12}=\frac{|U_{\rm PMNS}|_{12}}{\sqrt{1-(|U_{\rm PMNS}|_{13})^2}},\;s_{23}=\frac{|U_{\rm PMNS}|_{23}}{\sqrt{1-(|U_{\rm PMNS}|_{13})^2}},\nonumber\\
&&\delta={\rm arg}\Big(\frac{U_{\rm PMNS,12}U_{\rm PMNS,23}U_{\rm PMNS,13}^*U_{\rm PMNS,22}^*+s_{12}^2c_{13}^2s_{13}^2s_{23}^2}{c_{12}s_{12}c_{13}^2s_{13}c_{23}s_{23}}\Big),\nonumber\\
&&\rho={\rm arg}(U_{\rm PMNS,11}U_{\rm PMNS,13}^*)-\delta,\;\sigma={\rm arg}(U_{\rm PMNS,12}U_{\rm PMNS,13}^*)-\delta.\label{eq31}
\end{eqnarray}
In the context of the minimal seesaw mechanism, the Majorana CP phase associated with the massless neutrino has no physical impact. Consequently, only a single Majorana CP phase is physically relevant~\cite{Xing:2020ald}, i.e. $\sigma$ in the NH case ($m_{\nu_1}=0$) and $\sigma-\rho$ which can be redefined as $\sigma$~\cite{Mei:2003gn} in the IH case ($m_{\nu_3}=0$). In addition, to determine the best fit for the mass-squared differences in Eq.~(\ref{eq1}) and the flavor mixings, Dirac CP phase $\delta$ in Eq.~(\ref{eq3}), we perform a $\chi^2$ test, which can be constructed as
\begin{eqnarray}
&&\chi^2=\sum_1 \Big(\frac{O_i^{\rm th}-O_i^{\rm exp}}{\sigma_i^{\rm exp}}\Big)^2,
\end{eqnarray}
where $O_i^{\rm th}$ denotes the $i-$th observable computed theoretically, $O_i^{\rm exp}$ is the corresponding experimental value and $\sigma_i^{\rm exp}$ is the uncertainty in $O_i^{\rm exp}$.

\begin{figure}
\setlength{\unitlength}{1mm}
\includegraphics[width=2.6in]{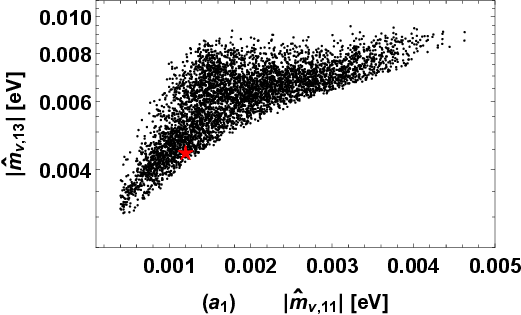}
\vspace{0.2cm}
\includegraphics[width=2.6in]{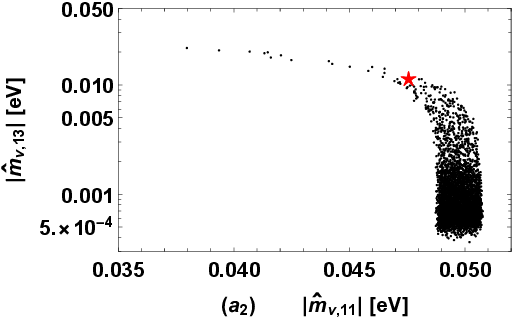}
\vspace{0.2cm}
\includegraphics[width=2.6in]{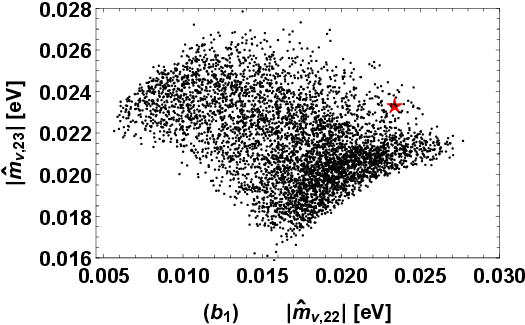}
\vspace{0.2cm}
\includegraphics[width=2.6in]{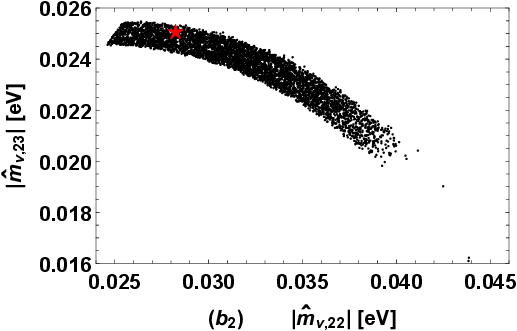}
\vspace{0.2cm}
\includegraphics[width=2.6in]{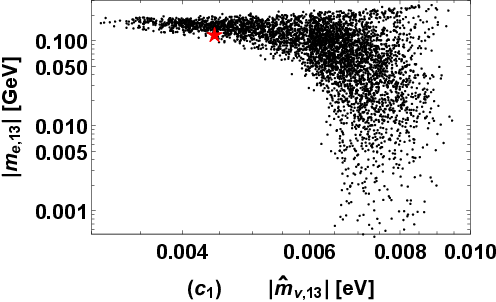}
\vspace{0.2cm}
\includegraphics[width=2.6in]{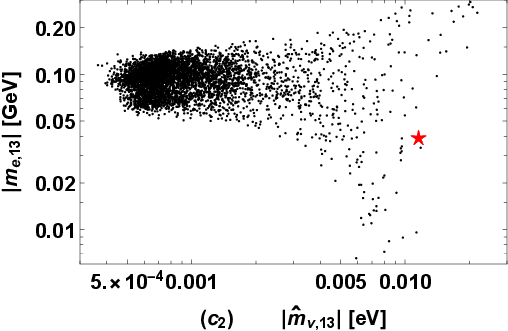}
\vspace{0.2cm}
\includegraphics[width=2.6in]{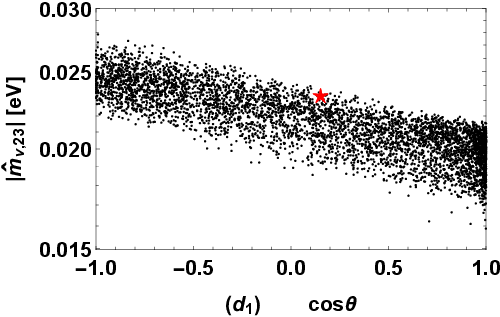}
\vspace{0.2cm}
\includegraphics[width=2.6in]{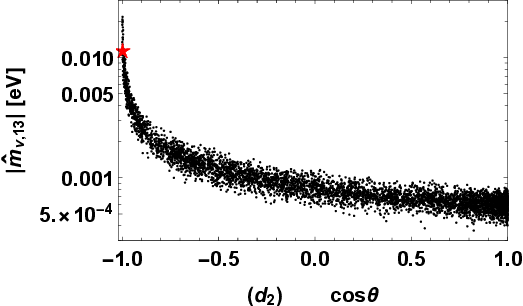}
\vspace{-0.8cm}
\caption[]{Scanning the parameter space in Eq.~(\ref{eq26}), and keeping the mass-squared differences, flavor mixings in the experimental $3\sigma$ ranges, the results of $|\hat m_{\nu,13}|-|\hat m_{\nu,11}|$ ($a_1$), $|\hat m_{\nu,23}|-|\hat m_{\nu,22}|$ ($b_1$), $|m_{e,13}|-|\hat m_{\nu,13}|$ ($c_1$), $|\hat m_{\nu,23}|-\cos\theta$ ($d_1$) for NH neutrino masses are plotted, where the red stars denote the best fit corresponding to $\chi^2=\mathbf{0.169}$. Similarly, the results for IH neutrino masses are plotted in ($a_2$, $b_2$, $c_2$, $d_2$), where the best fit corresponding to $\chi^2=\mathbf{0.312}$.}
\label{Fig2}
\end{figure}
Taking the charged lepton masses $m_e=0.511\;{\rm MeV},\;m_\mu=105.658\;{\rm MeV},\;m_\tau=1.777\;{\rm GeV}$ as inputs, we scan the parameter space
\begin{eqnarray}
&&|\hat m_{\nu,11}|\sim(10^{-6},\;10^{-1})\;{\rm eV},\;|\hat m_{\nu,22}|\sim(10^{-6},\;10^{-1})\;{\rm eV},\;|m_{e,13}|\sim(0,\;0.3)\;{\rm GeV},\nonumber\\
&&\theta_{\nu,\alpha\beta}\sim(-\pi,\;\pi),\;\theta_{e,ij}\sim(-\pi,\;\pi),\label{eq26}
\end{eqnarray}
where $\alpha\beta=11,22,13,23$, $ij=12,13,23$, $m_{e,33}$, $|m_{e,12}|$, $|m_{e,23}|$, $|\hat m_{\nu,13}|$, $|\hat m_{\nu,23}|$ can be obtained by Eqs.~(\ref{eq9}, \ref{eq15}, \ref{eq16}). Then keeping the mass-squared differences in Eq.~(\ref{eq1}) in the experimental $3\sigma$ ranges and Eq.~(\ref{eq25}) in the scanning, we plot the results of $|\hat m_{\nu,13}|-|\hat m_{\nu,11}|$, $|\hat m_{\nu,23}|-|\hat m_{\nu,22}|$, $|m_{e,13}|-|\hat m_{\nu,13}|$, $|\hat m_{\nu,23}|-\cos\theta$ for NH neutrino masses in Fig.~\ref{Fig2} ($a_1$), Fig.~\ref{Fig2} ($b_1$), Fig.~\ref{Fig2} ($c_1$), Fig.~\ref{Fig2} ($d_1$) respectively. Similarly, the results of $|\hat m_{\nu,13}|-|\hat m_{\nu,11}|$ ($a_2$), $|\hat m_{\nu,23}|-|\hat m_{\nu,22}|$ ($b_2$), $|m_{e,13}|-|\hat m_{\nu,13}|$ ($c_2$), $|\hat m_{\nu,13}|-\cos\theta$ ($d_2$) for IH neutrino masses are also plotted in Fig.~\ref{Fig2}. The red stars denote the best fit corresponding to $\chi^2=\mathbf{0.169}$ for NH neutrino masses and $\chi^2=\mathbf{0.312}$ for IH neutrino masses. The results of the best fit are listed in Tab.~\ref{tab1},
\begin{table*}
\begin{tabular*}{\textwidth}{@{\extracolsep{\fill}}llll@{}}
\hline
Observables & $O_i^{\rm th}$ & $O_i^{\rm exp}$ & Deviations in $\%$\\
\hline
NH\\
$\Delta m_{12}^2$[eV$^2$] & $7.56\times 10^{-5}$    & $7.53\times 10^{-5}$   & 0.40\\
$\Delta m_{23}^2$[eV$^2$] & $2.457\times 10^{-3}$   & $2.455\times 10^{-3}$  & 0.08\\
$s_{12}^2$                & 0.299                   & 0.307                  & 2.61\\
$s_{13}^2$                & $2.18\times 10^{-2}$    & $2.19\times 10^{-2}$   & 0.46\\
$s_{23}^2$                & 0.552                   & 0.558                  & 1.08\\
$\delta$                  & $1.16\pi$                    & $1.19\pi$                    & 2.52\\
\hline
IH\\
$\Delta m_{12}^2$[eV$^2$] & $7.40\times 10^{-5}$    & $7.53\times 10^{-5}$   & 1.73\\
$\Delta m_{32}^2$[eV$^2$] & $2.522\times 10^{-3}$   & $2.529\times 10^{-3}$  & 0.28\\
$s_{12}^2$                & 0.307                   & 0.307                  & 0.00\\
$s_{13}^2$                & $2.20\times 10^{-2}$    & $2.19\times 10^{-2}$   & 0.46\\
$s_{23}^2$                & 0.538                   & 0.553                  & 2.71\\
$\delta$                 & $1.22\pi$                    & $1.19\pi$                    & 2.52\\
\hline
\end{tabular*}
\caption{The results obtained for the best fit corresponding to $\chi^2=\mathbf{0.169}$ and $\chi^2=\mathbf{0.312}$ for NH and IH neutrino masses respectively.}
\label{tab1}
\end{table*}
where various $O_i^{\rm exp}$ are listed in the third column and $\sigma_i^{\rm exp}$ are take from Eqs.~(\ref{eq1}, \ref{eq3}). The parameters for the best fit listed in Tab.~\ref{tab1} are
\begin{eqnarray}
&&{\rm NH:}\;\hat m_{\nu,11}=0.00120e^{-1.46 i}\;{\rm eV},\;\hat m_{\nu,22}=0.0234e^{0.0665 i}\;{\rm eV},\;\hat m_{\nu,13}=0.00443e^{-0.678i}\;{\rm eV},\nonumber\\
&&\qquad\;\hat m_{\nu,23}=0.0234e^{2.52i}\;{\rm eV},\;m_{e,13}=0.12e^{2.94 i}\;{\rm GeV},\;\theta_{e,12}=-0.040,\;\theta_{e,23}=-1.30,\quad\nonumber\\
&&{\rm IH:}\;\hat m_{\nu,11}=0.0475e^{-1.00 i}\;{\rm eV},\;\hat m_{\nu,22}=0.0282e^{2.87 i}\;{\rm eV},\;\hat m_{\nu,13}=0.0116e^{-2.27i}\;{\rm eV},\nonumber\\
&&\qquad\;\hat m_{\nu,23}=0.0251e^{-1.86i}\;{\rm eV},\;m_{e,13}=0.0399e^{-2.98 i}\;{\rm GeV},\;\theta_{e,12}=-1.34,\;\theta_{e,23}=-0.397.\nonumber\\ \label{eq28}
\end{eqnarray}

Fig.~\ref{Fig2} and Tab.~\ref{tab1} clearly demonstrate that for both NH and IH neutrino masses, the lepton sector in the FDM can fit the measured charged lepton masses, the mass-squared differences of neutrinos and flavor mixings described by the PMNS matrix well. The preferred parameter spaces for NH and IH neutrino masses are notably different, as can be observed by comparing Fig.~\ref{Fig2} ($a_1$, $b_1$, $c_1$, $d_1$) with Fig.~\ref{Fig2} ($a_2$, $b_2$, $c_2$, $d_2$). To match the measured mass-squared differences of neutrinos, the values of $|\hat m_{\nu,13}|$, $|\hat m_{\nu,23}|$ are related closely to the selected values of $|\hat m_{\nu,11}|$, $|\hat m_{\nu,22}|$, $\cos\theta$ as shown in Fig.~\ref{Fig2} ($a_1$, $a_2$, $b_1$, $b_2$, $d_1$, $d_2$), and the fact can be easily understood from Eqs.~(\ref{eq15}, \ref{eq16}). Additionally, to fit the measured PMNS matrix elements, the value of $|m_{e,13}|$ is also related to the chosen values of $|\hat m_{\nu,13}|$.

\begin{figure}
\setlength{\unitlength}{1mm}
\includegraphics[width=2.6in]{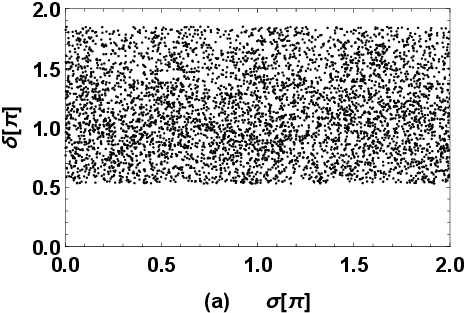}
\vspace{0.2cm}
\includegraphics[width=2.6in]{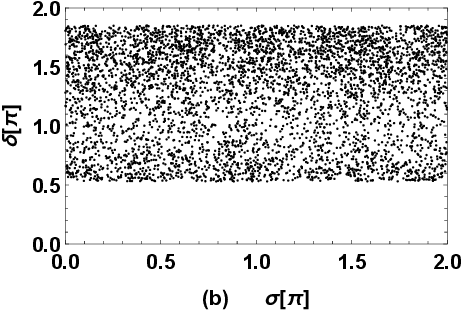}
\vspace{-0.0cm}
\caption[]{Based on the results obtained in Fig.~\ref{Fig2}, the allowed range of $\delta-\sigma$ is plotted, where (a) for NH neutrino masses and (b) for IH neutrino masses.}
\label{Figadd}
\end{figure}
\begin{figure}
\setlength{\unitlength}{1mm}
\includegraphics[width=2.6in]{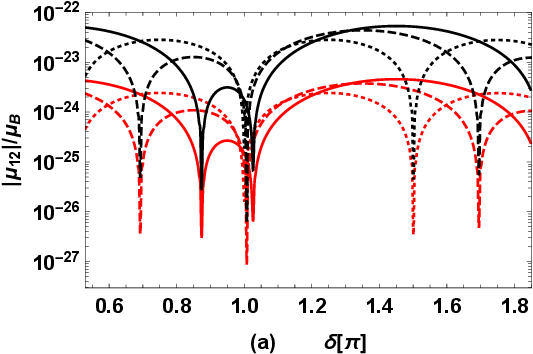}
\vspace{0.2cm}
\includegraphics[width=2.6in]{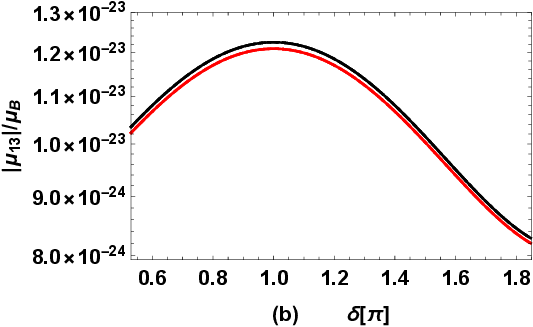}
\vspace{0.2cm}
\includegraphics[width=2.6in]{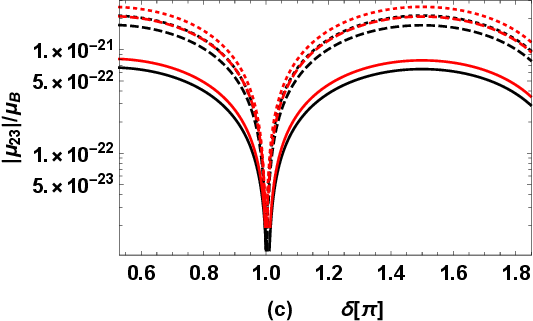}
\vspace{0.2cm}
\includegraphics[width=2.6in]{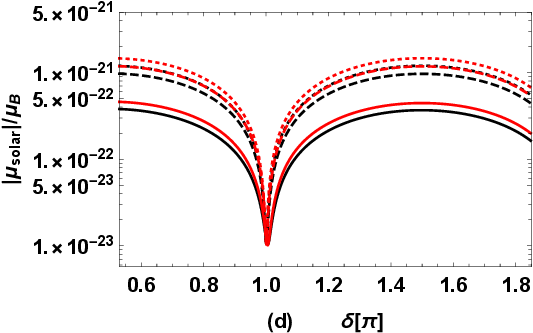}
\vspace{-0.8cm}
\caption[]{The results of $|\mu_{12}|$ (a), $|\mu_{13}|$ (b), $|\mu_{23}|$ (c), $|\mu_{\rm solar}|$ (d) versus $\delta$ are plotted. The black curves denote the results for NH neutrino masses, the red curves denote the results for IH neutrino masses, and the solid, dashed, dotted curves denote the results for $\sigma=0.1\pi,\;0.3\pi,\;0.5\pi$ respectively.}
\label{Fig3}
\end{figure}

As discussed earlier, observing the transition MDM of neutrinos experimentally is one of the most effective methods for providing definitive evidence of neutrino-related NP, particularly the Dirac and Majorana neutrinos have quite different structures in their MDM. To investigate the effects of Dirac CP phase $\delta$ and Majorana CP phase $\sigma$ on the transition MDM of neutrinos, we present the allowed range of $\delta-\sigma$ based on the results obtained in Fig.~\ref{Fig2}, where Fig.~\ref{Figadd} (a) for NH neutrino masses and Fig.~\ref{Figadd} (b) for IH neutrino masses. The picture shows obviously that the ranges $0.53\pi\leq\delta\leq1.85\pi$ ($3\sigma$ ranges provided by PDG) and $0\leq\sigma\leq2\pi$ are allowed. Then, the results of $|\mu_{12}|$, $|\mu_{13}|$, $|\mu_{23}|$, $|\mu_{\rm solar}|$ versus $\delta$ are plotted in Fig.~\ref{Fig3} (a), (b), (c), (d) respectively. In these plots, the black curves represent the results for NH neutrino masses, the red curves represent the results for IH neutrino masses, and the solid, dashed, dotted curves correspond to $\sigma=0.1\pi,\;0.3\pi,\;0.5\pi$ respectively.

Fig.~\ref{Fig3} illustrates that the Dirac CP phase $\delta$ and the Majorana CP phase $\sigma$, significantly affect $|\mu_{12}|$, $|\mu_{13}|$, $|\mu_{23}|$, $|\mu_{\rm solar}|$. However, $|\mu_{13}|$ is nearly independent of $\sigma$ as shown in Fig.~\ref{Fig3} (b). The comparison of the black and red curves in Fig.~\ref{Fig3} suggests that precise measurements of $|\mu_{12}|$, $|\mu_{13}|$, $|\mu_{23}|$, $|\mu_{\rm solar}|$ could help determine the hierarchy of neutrino masses. For both NH and IH neutrino masses, $|\mu_{12}|$, $|\mu_{23}|$, $|\mu_{\rm solar}|$ become extremely small when $\delta\approx\pi$ while $|\mu_{13}|$ does not exhibit this behavior. Moreover, the contributions from $\sigma$ to $|\mu_{12}|$ can cancel those from $\delta$ by selecting appropriate values for $\sigma$ and $\delta$, resulting in the presence of additional valleys in Fig.~\ref{Fig3} (a). It is also noteworthy that $|\mu_{12}|$, $|\mu_{13}|$, $|\mu_{23}|$ can reach approximately $5\times10^{-23}\mu_B$, $1.2\times10^{-23}\mu_B$, $3\times10^{-21}\mu_B$, respectively. Observations of supernova fluxes in the JUNO experiment may reveal the effects of collective spin-flavour oscillations due to the Majorana neutrino transition MDM about $10^{-21}\mu_B$~\cite{JUNO:2015zny,Giunti:2015gga,Lu:2016ipr}. Studies of Majorana neutrino transition MDMs in the context of supernova explosions~\cite{deGouvea:2012hg,deGouvea:2013zp} indicate that these moments may leave a potentially observable imprint on the energy spectra of neutrinos and antineutrinos from supernovae, even for the moments as small as $10^{-24}\mu_B$. Other new possibilities for neutrino transition MDMs visualization in extreme astrophysical environments are considered recently in Refs.~\cite{Grigoriev:2017wff,Kurashvili:2017zab}. These suggest that the transition MDMs of Majorana neutrinos predicted in the FDM have a potential to be observed. Moreover, once the Dirac CP phase $\delta$ is measured precisely, the Majorana CP phase $\sigma$ can be determined through the observations of neutrino transition MDMs. This is because the predicted neutrino transition MDMs in the framework of FDM are dominated by $\delta$, $\sigma$ and the lightest neutrino mass (which is predicted to be $0$ in the FDM). In this case, all of the free parameters in Eq.~(\ref{eq10}), Eq.~(\ref{eq11}) are determined by charged lepton mass, neutrino masses, PMNS matrix including Dirac CP phase $\delta$ and Majorana CP phase $\sigma$. It indicates that the parameter space of the FDM is much more tightly constrained, and the predictive power of the FDM in the lepton sector is enhanced greatly.

\section{Summary\label{sec4}}

Fitting the neutrino-related measurements theoretically is important for exploring NP theories beyond the SM. Motivated by the flavor-related mass matrices of neutrinos and charged leptons in the FDM, our focus is on the neutrino sector in the model. Analytical calculations of the neutrino masses reveal that the lightest neutrino is predicted to be massless in the FDM, with the masses of the other two neutrinos being determined by the measured neutrino mass-squared differences. And determining the neutrino masses is one of the most important purposes of the next generation neutrino-related experiments. The numerical results demonstrate that, for both NH and IH neutrino masses, the lepton sector in the FDM can accurately fit the measured charged lepton masses, the neutrino mass-squared differences, the flavor mixing angles and the Dirac CP phase. Furthermore, the dominant contributions to the neutrino transition MDM in the model come from the $W$ boson-related interactions. The Dirac CP phase $\delta$ and the Majorana CP phase $\sigma$ significantly influence the theoretical predictions of these neutrino transition MDM. It is also noteworthy that the predicted values of $|\mu_{12}|$, $|\mu_{13}|$, $|\mu_{23}|$, $|\mu_{\rm solar}|$ can reach about $5\times10^{-23}\mu_B$, $1.2\times10^{-23}\mu_B$, $3\times10^{-21}\mu_B$, $4\times10^{-21}\mu_B$ respectively, which may have the potential to be observed in the JUNO experiment and the energy spectra of neutrinos from supernovae. Once the Dirac CP phase $\delta$ and neutrino transition MDMs are measured precisely, all of the free parameters in Eq.~(\ref{eq10}), Eq.~(\ref{eq11}) are determined by charged lepton mass, neutrino masses, PMNS matrix including Dirac CP phase $\delta$ and Majorana CP phase$\sigma$, which greatly enhances the predictive power of the FDM in the lepton sector.

\begin{acknowledgments}
The work has been supported by the National Natural Science Foundation of China (NNSFC) with Grants No. 12075074, No. 12235008, Hebei Natural Science Foundation for Distinguished Young Scholars with Grant No. A2022201017, No. A2023201041, Natural Science Foundation of Guangxi Autonomous Region with Grant No. 2022GXNSFDA035068, and the youth top-notch talent support program of the Hebei Province.

\end{acknowledgments}

\end{CJK*}

\end{document}